\documentclass[twocolumn,showpacs,preprintnumbers,amsmath,amssymb,aps]{revtex4-1}
\usepackage{graphicx}% Include figure files
\usepackage{dcolumn}% Align table columns on decimal point,
\usepackage{bm}% bold math
\usepackage{float}
\usepackage{color}
\usepackage{soul}
\usepackage{ragged2e}
\begin{document}

\title{Duality in quantum work}
\author{Bao-Ming Xu$^{1,2}$}
\email{xbmv@bit.edu.cn}
\author{Zhan Chun Tu$^{2}$}
\email{tuzc@bnu.edu.cn}
\author{Jian Zou$^{3}$}
\email{zoujian@bit.edu.cn}
\affiliation{$^{1}$Shandong Key Laboratory of Biophysics, Institute of Biophysics, Dezhou University, Dezhou 253023, China}%
\affiliation{$^{2}$Department of Physics, Beijing Normal University, Beijing 100875, China}
\affiliation{$^{3}$School of Physics, Beijing Institute of Technology, Beijing 100081, China}

\date{Submitted \today}

\begin{abstract}
An open question of fundamental importance in quantum thermodynamics is how to describe the statistics of work for the initial state with quantum coherence. In this paper, work statistics is considered from a fully new perspective of ``wave-particle" duality. Based on the generalized quantum work measurement, predictability of energy levels $\mathcal{D}_W$ and effectiveness of coherence $\mathcal{V}_W$ are defined, and they obey inequality $\mathcal{D}_W^2+\mathcal{V}_W^2\leq1$, which is the fundamental tradeoff relations between the contributions of population and coherence to quantum work distribution. As an application, we consider a driven two-level system and discuss the condition of the bound of above tradeoff relation. These results shed light on the effects of quantum coherence in quantum thermodynamics.
\end{abstract}

\maketitle

\section{Introduction}
Motivated by recent experimental progress in fabrication and manipulation of micro and nanoscale objects \cite{Bloch2008,Kippenberg2007,Aspelmeyer2014}, there is an urgent need of a theoretical foundation of what thermodynamics quantity means in quantum mechanics and how to extend the principles of thermodynamics to the quantum domain. In the quantum regime, it is very hard to define work because it is not an observable \cite{Talkner2007a}. Traditionally, quantum work is defined as the difference of the energies obtained by two-point measurement scheme (TPM): performing two projective energy measurements at the beginning and the end of external protocol. Based on TPM, the extension of classical fluctuation theorems \cite{Jarzynski1997,Crooks1999,Seifert2005,Esposito2010,Ueda2010,Harris2007,Seifert2012,Jarzynski2004} to the quantum regime is obtained \cite{Jarzynski2004,Tasaki2000,Kurchan2000,Mukamel2003,Talkner2007a,Talkner2007b,Talkner2009,Campisi2009,Esposito2009,Crooks2008,Andrieux2009,Campisi2011,Watanabe2014}, see reviews \cite{Esposito2009,Campisi2011} for detail discussion, and these fluctuation theorems have been experimentally verified in various systems \cite{Batalho2014,Huber2008,An2015,Hoang2018,Xiong2018,Cerisola2017,Naghiloo2018}. However, if the system is initially in the superposition of some energy levels, quantum coherence will be completely destroyed by the first measurement, and the work fluctuation relation is not ``quantum" to some extent. It should be noted that due to the destruction of quantum coherence, the first law of thermodynamics can not be satisfied.

In order to include the effects of the initial quantum coherence, projective measurement was replaced by Gaussian measurement \cite{Watanabe2014,Talkner2016,Solinas2017}, and a modified Jarzynski equality was obtained \cite{Watanabe2014}. Besides two-measurement scheme, some single-measurement schemes were proposed \cite{Roncaglia2014,Allahverdyan2014,Chiara2015,Solinas2015,Solinas2016}. Some of these results satisfy the first law of thermodynamics but can not recover fluctuation theorems in thermal equilibrium limit \cite{Roncaglia2014,Allahverdyan2014,Chiara2015}, and the others can simultaneously satisfy the first law of thermodynamics and fluctuation theorems, but negative probability appears \cite{Solinas2015,Solinas2016,Xu2018}. Other efforts based on Bohmian framework \cite{Sampaio2018}, autonomous framework \cite{Aberg2018,Holmes2018}, quantum feedback control \cite{Alonso2016}, etc., for including the effects of quantum coherence have also been made, and also fluctuation theorems can not be recovered. Recently, Llobet \textit{et al}. proved that for the initial state with quantum coherence, the first law of thermodynamics and fluctuation theorem can not be simultaneously satisfied \cite{Llobet2017}. Understanding the interplay between the contributions of population and quantum coherence to work statistics is a pressing problem in quantum thermodynamics.

Population manifests which-level information, and coherence can give rise to some interference effects. This reminds us ``wave-particle" duality \cite{Englert1996} which describes the uncertainty relation between interference pattern and acquisition of which-way information, and has been widely investigated in many aspects \cite{Durr1998,Peter1999,Jacques2008,Bera2015,Bagan2016,Coles2016,Bagan2018,Qureshi2017}. In this paper, we consider work distribution from the ``wave-particle" duality perspective, and aim at giving a quantitative fundamental relation between the contributions of population and coherence to quantum work distribution.

This paper is organized as follows: In next section we discuss the duality in quantum work distribution. As an application we consider a driven two-level system in Sec. III. Finally, Sec. IV closes the paper with some concluding remarks.

\section{Duality in quantum work}

Consider a closed quantum system described by a Hamiltonian $\hat{H}_s(\lambda_t)$ that depends on an externally controlled parameter $\lambda_t$ changed from $\lambda_0$ to $\lambda_{t'}$. The Hamiltonian has a spectral decomposition $\hat{H}_s(\lambda_t)=\sum_{n}E^t_n|E^t_n\rangle\langle E^t_n|$. Externally controlled evolution of the system is described by a unitary operator $U_s(t')\equiv\overleftarrow{T}\exp{\{-i\int_{0}^{t'}\hat{H}_s(\lambda_t)dt\}}$ ($\overleftarrow{T}$ is the time ordering operator). Traditionally, work done by external control protocol is determined by TPM: At the beginning, the first projective measurement onto the eigenbasis of the initial Hamiltonian $\hat{H}_s(\lambda_0)$ is performed; then, the system evolves under unitary dynamics $U_s(t')$ generated by protocol $\lambda_0\rightarrow\lambda_{t'}$; finally, the second projective measurement onto the eigenbasis of the final Hamiltonian $\hat{H}_s(\lambda_{t'})$ is performed. One should prepare many copies of the system with the same state $\rho$, and then perform TPM for each copy. For each trial, one may obtain $E^0_n$ for the first measurement outcome followed by $E^{t'}_m$ for the second measurement, and the corresponding probability is $P_n^0P^{t'}_{m|n}$, where $P_n^0=\langle E^0_n|\rho|E^0_n\rangle$ and $P^{t'}_{m|n}=|\langle E^{t'}_m|U_s(t')|E^0_n\rangle|^2$. The trajectory work is defined as $W=E^{t'}_m-E^{0}_n$ and whose probability distribution is $P(W)=\sum_{mn}P_n^0P^{t'}_{m|n}\delta(W-(E^{t'}_m-E^{0}_n))$. However, the destruction of quantum coherence (if it exists) is the main drawback of TPM, which makes the application of TPM to investigate the effects of quantum coherence impossible. To investigate the contribution of quantum coherence to work statistics is the task of the present paper.

In this paper, we adopt a positive operator-valued measure (POVM) to estimate work statistics. A POVM is a set of non-negative Hermitian operators $\{\mathcal{M}^W\}$, which satisfy $\int\mathcal{M}^WdW=\mathbb{I}$ with $\mathbb{I}$ being the identity matrix. Each possible value of work $W$ is associated with an operator $\mathcal{M}^W$. The probability to obtain $W$ can be calculated through the generalized Born rule:
\begin{equation}\label{Born rule}
\mathcal{P}(W)=\mathrm{Tr}[\mathcal{M}^W\rho].
\end{equation}
The element of POVM $\mathcal{M}^W$ depends on the process, $\Pi=(\hat{H}_s(\lambda_0),\hat{H}_s(\lambda_{t'}),U_s(t'))$, but in order to be a universal scheme it must be independent of the initial state $\rho$ \cite{Llobet2017}. For the POVM measurement, the post-measurement state of the system is of little interest, which is very different from TPM, wherein the system state collapses to the eigenbasis of $\hat{H}_s(\lambda_{0})$ and $\hat{H}_s(\lambda_{t'})$ after the first and the second projective measurements. For the traditional TPM, the corresponding element of POVM is $\mathcal{M}^W=\sum_{mn}P^{t'}_{m|n}|E^0_n\rangle\langle E^0_n|\delta(W-(E^{t'}_m-E^0_n))$.

In the eigenbasis of the initial Hamiltonian $\hat{H}_s(\lambda_0)$, an arbitrary initial state of the system can be expressed as
\begin{equation}\label{}
  \rho=\sum_{ij}\rho_{ij}|E^0_i\rangle\langle E^0_j|
\end{equation}
with $\rho_{ij}\equiv\langle E^0_i|\rho|E^0_j\rangle$. In fact, diagonal element $\rho_{ii}$ is the probability of obtaining $E^0_i$ by the first measurement in TPM, i.e, $\rho_{ii}=P_i^0$. In this paper, we divide the initial state $\rho$ into two parts:
\begin{equation}\label{}
  \rho=\rho_{in}+\rho_c
\end{equation}
where
\begin{equation}\label{}
  \rho_{in}=\sum_i\rho_{ii}|E^0_i\rangle\langle E^0_i|
\end{equation}
is the incoherent part and
\begin{equation}\label{}
  \rho_{c}=\sum_i\sum_{j\neq i}\rho_{ij}|E^0_i\rangle\langle E^0_j|
\end{equation}
is the coherent part. Accordingly, quantum work distribution (\ref{Born rule}) can be divided into two parts:
\begin{equation}\label{}
 \mathcal{P}(W)= \mathcal{P}_{in}(W)+ \mathcal{P}_{c}(W),
\end{equation}
where
\begin{equation}\label{}
\begin{split}
  &\mathcal{P}_{in}(W)\equiv\mathrm{Tr}[\mathcal{M}^W\rho_{in}]=\sum_i\rho_{ii}\langle E^0_i|\mathcal{M}^W|E^0_i\rangle
\end{split}
\end{equation}
and
\begin{equation}\label{}
\begin{split}
  \mathcal{P}_{c}(W)&\equiv\mathrm{Tr}[\mathcal{M}^W\rho_{c}]=\sum_i\sum_{j\neq i}\rho_{ij}\langle E^0_i|\mathcal{M}^W|E^0_j\rangle
\end{split}
\end{equation}
are respectively the incoherent work distribution and the coherent work distribution we named. In fact, $\langle E^0_{i}|\mathcal{M}^W|E^0_{i}\rangle$ is quantum work distribution when the system initially stays at a definite energy level $|E^0_{i}\rangle$, i.e.,
\begin{equation}\label{}
  \mathcal{P}_i(W)=\langle E^0_{i}|\mathcal{M}^W|E^0_{i}\rangle.
\end{equation}
Physically, $\rho_{in}$ and $\rho_{c}$ are essentially the population and the coherence of the energy levels of the system before external protocol. Thus incoherent and coherent work distributions $\mathcal{P}_{in}(W)$ and $\mathcal{P}_{c}(W)$ record the information of population and coherence before external protocol. Now we would ask how much information of population and coherence can be manifested by quantum work distribution, or how much information of population and coherence contributes to quantum work distribution.

The ``wave-particle" duality describes the exclusion of interference pattern and the acquisition of which-way information \cite{Englert1996}. Now we consider quantum work distribution $\mathcal{P}(W)$ from the ``wave-particle" duality perspective. In this perspective, energy levels can be considered as the ``particle" like nature and their coherence can be considered as the ``wave" like nature. Because quantum work distribution records the information of population and coherence, quantum work can behave both the ``particle" like nature and the ``wave" like nature, called ``trajectory-coherence" duality. Just like the double slit experiment: If only a single slit is open, i.e., the trajectory of particle is known, interference fringes can not be observed on screen; if two slits are both open, i.e., the trajectory of particle is unknown, interference fringes can be observed on screen. So, the information of population and the information of coherence can not be simultaneously manifested by quantum work distribution. Then it is nature to ask how to quantitatively describe this incompatibility?

In order to quantify the contribution of population, we define the predictability of energy levels as
\begin{equation}\label{predictability}
  \mathcal{D}_W=\frac{1}{2(d-1)}\sum_{mn}\int dW\Big\vert \rho_{mm}\mathcal{P}_m(W)-\rho_{nn}\mathcal{P}_n(W)\Big\vert,
\end{equation}
where, $d$ is the dimension of the density matrix of the system. $\mathcal{D}_W$ quantifies the distinguishability of energy levels, i.e., the amount of which-level information available by quantum work distribution. According to $|A-B|\leq|A|+|B|$,
%$\sum_{mn}\int dW|\rho_{mm}\mathcal{P}_m(W)-\rho_{nn}\mathcal{P}_n(W)|/2\leq\sum_m\sum_{n>m}\int dW(\rho_{mm}\mathcal{P}_m(W)+\rho_{nn}\mathcal{P}_n(W))=d-1$, and so
it can be proved that
\begin{equation}\label{}
  \mathcal{D}_W\leq1.
\end{equation}
If $\mathcal{D}_W=1$, $\mathcal{P}_m(W)=0$ or $\mathcal{P}_n(W)=0$, which means that there is no overlap between two quantum work distributions of $\mathcal{P}_m(W)$ and $\mathcal{P}_n(W)$, and energy levels can be completely distinguished by quantum work distribution. If, on the other hand, $\mathcal{D}_W=0$, $\mathcal{P}_m(W)=\mathcal{P}_n(W)$ for any work value $W$, so that energy levels cannot be distinguished at all.

%The boundary of $|\rho_{mm}\mathcal{P}_m(W)-\rho_{nn}\mathcal{P}_n(W)|\leq\rho_{mm}\mathcal{P}_m(W)+\rho_{nn}\mathcal{P}_n(W)$ is found for

We also define the effectiveness of quantum coherence to quantify the contribution of quantum coherence:
\begin{equation}\label{visibility}
  \mathcal{V}_W=\frac{1}{d-1}\int dW\Big\vert\mathcal{P}_c(W)\Big\vert.
\end{equation}
Because $|\sum\cdot|\leq\sum|\cdot|$, $|\mathcal{P}_c(W)|\leq\sum_{n}\sum_{m\neq n}|\rho_{mn}||\langle E^0_n|\mathcal{M}^W|E^0_m\rangle|$. According to the non-negativity of $\mathcal{M}^W$, i.e., $\langle E^0_n|\mathcal{M}^W|E^0_m\rangle\langle E^0_m|\mathcal{M}^W|E^0_n\rangle\leq \langle E^0_m|\mathcal{M}^W|E^0_m\rangle\langle E^0_n|\mathcal{M}^W|E^0_n\rangle$, we can prove that
\begin{equation}\label{}
  \mathcal{V}_W\leq\frac{\mathcal{C}}{d-1}
\end{equation}
%$\mathcal{V}_W\leq\int dW\sum_{n,m\neq n}|\rho_{mn}|\sqrt{\mathcal{P}_{m}(W)\mathcal{P}_{n}(W)}/(d-1)\leq\sum_{n,m\neq n}|\rho_{mn}|\int dW(\mathcal{P}_m(W)+\mathcal{P}_n(W))/[2(d-1)]=\mathcal{C}/(d-1)$,
with $\mathcal{C}\equiv\sum_{n,m\neq n}|\rho_{mn}|$ being the degree of quantum coherence \cite{coherence}. If the initial state $\rho$ is maximally coherent, $\mathcal{C}=\sum_{n,m\neq n}|\rho_{mn}|=\sum_{n,m\neq n}(\rho_{mm}+\rho_{nn})/2=d-1$, so
\begin{equation}\label{}
  \mathcal{V}_W\leq1.
\end{equation}
Given an initial state $\rho$, we can choose a proper work measurement $\mathcal{M}^W$ to maximize $\mathcal{V}_W$. If $\mathcal{V}_W$ is maximized, i.e., $\mathcal{V}_W=\mathcal{C}/(d-1)$, quantum coherence will be completely manifested. In order to further understand the physical meaning of $\mathcal{V}_W$, we introduce the trace distance $D(\mathcal{P}(W),\mathcal{P}_{in}(W))\equiv\int dW|\mathcal{P}(W)-\mathcal{P}_{in}(W)|/2=\int dW|\mathcal{P}_{c}(W)|/2$. The effectiveness of quantum coherence $\mathcal{V}_W$ is essentially equivalent to this trace distance, i.e., $\mathcal{V}_W=2D(\mathcal{P}(W),\mathcal{P}_{in}(W))/(d-1)$. In other words, the contribution of quantum coherence is manifested by the trace distance between quantum work distribution $\mathcal{P}(W)$ and incoherent work distribution $\mathcal{P}_{in}(W)$. If $\mathcal{V}_W$ is maximized, $D(\mathcal{P}(W),\mathcal{P}_{in}(W))=\mathcal{C}/2$, quantum coherence of $\rho$ can be completely acquired by the trace distance $D(\mathcal{P}(W),\mathcal{P}_{in}(W))$. On the contrary, if $\mathcal{V}_W=0$, $D(\mathcal{P}(W)|\mathcal{P}_{in}(W))=0$, for example $\mathcal{P}(W)$ and $\mathcal{P}_{in}(W)$ are the same in TPM, the quantum coherence of $\rho$ can not be observed at all through quantum work distribution.

The exclusion between energy level predictability and quantum coherence effectiveness implies the central result of this paper that the predictability $\mathcal{D}_W$ and the effectiveness of coherence $\mathcal{V}_W$ obey the inequality
\begin{equation}\label{DV2_inequality}
  \mathcal{D}_W^2+\mathcal{V}_W^2\leq1,
\end{equation}
which is a fundamental quantitative statement about duality in the quantum work distribution. In particular, the extreme situations characterized by perfect effectiveness of coherence or full which-level information are mutually exclusive.

Now we give the proof of the central results of this paper Eq. (\ref{DV2_inequality}). According to the definition of integral $\int f(W)dW=\sum_if(W_i)\Delta W$, the predictability of levels can be written as $\mathcal{D}_W=\sum_{i}\sum_{n,m\neq n}|\rho_{mm}\mathcal{P}_m(W_i)\Delta W-\rho_{nn}\mathcal{P}_n(W_i)\Delta W|/[2(d-1)]=\sum_{i}\sum_{n,m\neq n}v_{imn}[1-|u_{imn}|^2]^{1/2}/(d-1)$ with $v_{imn}=[\rho_{mm}\mathcal{P}_m(W_i)\Delta W+\rho_{nn}\mathcal{P}_n(W_i)\Delta W]/2$ and $|u_{imn}|=[\rho_{mm}\mathcal{P}_m(W_i)\Delta W\rho_{nn}\mathcal{P}_n(W_i)\Delta W]^{1/2}/v_{imn}$. For simplicity, we rewrite $\sum_{i}\sum_{n,m\neq n}$ as $\sum_k$, where $k$ takes all the possibilities of the set of $\{i,n,m\neq n\}$. In this simple case, $\mathcal{D}_W=\sum_kv_k[1-|u_k|^2]^{1/2}/(d-1)$. It can be seen that $v_k\geq0$, $\sum_kv_k\leq1$ and $|u_k|\leq1$. The effectiveness of coherence $\mathcal{V}_W=\sum_{i}\sum_{n,m\neq n}|\rho_{mn}(0)\mathcal{M}^{W_i}_{nm}\Delta W|/(d-1)\leq\sum_{i}\sum_{n,m\neq n}[\rho_{mm}\rho_{nn}\mathcal{P}_m(W_i)\Delta W\mathcal{P}_n(W_i)\Delta W]^{1/2}/(d-1)=\sum_kv_k|u_k|/(d-1)$. $(d-1)^2[\mathcal{D}_W^2+\mathcal{V}_W^2]\leq\sum_{kk'}v_kv_{k'}[(1-|u_k|^2)^{1/2}(1-|u_{k'}|^2)^{1/2}+|u_ku_{k'}|]$. Because $|u_k|\leq1$, $[\cdots]\leq1$ holds for the square brackets in the above inequality \cite{Englert1996}, therefore $(d-1)^2[\mathcal{D}_W^2+\mathcal{V}_W^2]\leq\sum_{kk'}v_kv_{k'}=(d-1)^2$, i.e., $\mathcal{D}_W^2+\mathcal{V}_W^2\leq1$.

\begin{figure*}
\centering
\includegraphics[width=16cm]{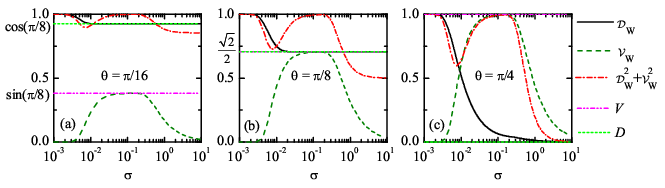}
\parbox{17.9cm}{\textbf{Fig. 1}. \justifying (Color online) $\mathcal{D}_W$ (black solid), $\mathcal{V}_W$ (olive dashed) and $\mathcal{D}_W^2+\mathcal{V}_W^2$ (red dashed dotted) as a function of $\sigma$ for (a) $\theta=\pi/16$, (b) $\theta=\pi/8$ and (c) $\theta=\pi/4$. The energy level predictability and the quantum coherence effectiveness of initial state are $D=\cos2\theta$ and $V=\sin2\theta$ respectively, which are plotted by using green short dashed and magenta dashed dotted dotted lines. For panel (b), $D$ and $V$ are coincide with each other. For all the panels, $\omega_0=0.01$, $\omega=\omega_0$ and $t=100$.}
\end{figure*}

\section{An Example: Driven Two-Level System}
As an application, in this section we consider a two-level system driven by a time dependent field and investigate the duality from its work distribution. Our main aim is to further understand the physical meaning of $\mathcal{D}_W$ and $\mathcal{V}_W$ and give the conditions of the bound of fundamental inequality (\ref{DV2_inequality}) by this simple model. As shown in the following, this is an instructive discussion because it gives some general results that is independent of the model. The Hamiltonian of driven two-level system is
\begin{equation}\label{Hamiltonian}
  H(\tau)=\omega_0\sigma_z+g\sin\omega\tau\sigma_x,
\end{equation}
where $\sigma_z=|2\rangle\langle2|-|1\rangle\langle1|$ and $\sigma_x=|2\rangle\langle1|+|1\rangle\langle2|$ are Pauli operators. $\omega_0$ is the frequency of the system, $\omega$ and $g$ are the frequency and the strength of driving. In this paper we consider $g$ as the energy scale, i.e., $g=1$. The transient eigenvalues of Hamiltonian (\ref{Hamiltonian}) are
\begin{equation}\label{}
\begin{split}
  \varepsilon_{1}^\tau=-\sqrt{\omega_0^2+\sin^2\omega\tau},~~\varepsilon_{2}^\tau=\sqrt{\omega_0^2+\sin^2\omega\tau}
\end{split}
\end{equation}
and the corresponding eigenvectors are
\begin{equation}\label{}
\begin{split}
&|\varepsilon_{1}^\tau\rangle=\frac{1}{\sqrt{(\omega_0-\varepsilon^\tau_1)^2+\sin^2\omega\tau}}
\biggl[(\omega_0-\varepsilon^\tau_1)|1\rangle-\sin\omega\tau|2\rangle\biggr] \\
&|\varepsilon_{2}^\tau\rangle=\frac{1}{\sqrt{(\omega_0+\varepsilon^\tau_2)^2+\sin^2\omega\tau}}
\biggl[(\omega_0+\varepsilon^\tau_2)|2\rangle+\sin\omega\tau|1\rangle\biggr]
\end{split}
\end{equation}
respectively.

To define the work of driving governed by $U(t)=\overleftarrow{T}\exp[\int_0^{t}-i H(\tau)d\tau]$, we need to know the initial and the final system energies. Traditionally, energy is obtained by TPM in which quantum coherence is destroyed. In order to include the effects of quantum coherence, we adopt the minimally disturbing energy measurement: The energy measurements at the beginning and the end of external protocol are performed by Gaussian superpositions of projective measurements of energy eigenstates, which can be expressed as \cite{Watanabe2014}
\begin{equation}\label{}
 M_{E^\tau}=\sum_{n=1}^2\frac{1}{(2\pi\sigma^2)^{1/4}}\exp\Bigl\{-\frac{(\varepsilon_n^\tau-E^\tau)^2}{4\sigma^2}\Bigr\}|\varepsilon_n^\tau\rangle\langle \varepsilon_n^\tau|,
\end{equation}
where $\sigma$ is the measurement error. Work is defined as the difference between the final and initial energies, i.e., $W=E^{t}-E^0$, whose distribution is
\begin{equation}\label{}
  \mathcal{P}(W)=\int dE^0\int dE^{t}\delta[W-(E^{t}-E^0)]P(E^{t},E^0),
\end{equation}
where
\begin{equation}\label{}
  P(E^{t},E^0)=\mathrm{Tr}\Bigl[M_{E^{t}}U(t)M_{E^0}\rho(0)M_{E^0}^\dag U^\dag(t)M_{E^{t}}^\dag\Bigr]
\end{equation}
is the joint probability of getting $E^{t}$ and $E^0$. In this minimally disturbing energy measurement scheme, the POVM operator is
\begin{equation}\label{}
  \mathcal{M}^W=\small\int dE^0\int dE^{t}\delta[W-(E^{t}-E^0)]M_{E^0}^\dag U^\dag(t)M_{E^{t}}^2U(t)M_{E^0}.
\end{equation}
The system is initially prepared in the superposition of energy levels, i.e., $\rho=|\Psi\rangle\langle\Psi|$ with $|\Psi\rangle=\sin\theta|2\rangle+\cos\theta|1\rangle$.

According to Eq. (\ref{predictability}), the predictability of levels is
\begin{equation}\label{DWE}
\begin{split}
\mathcal{D}_W=&\int dW\Big|\cos^2\theta\big|\langle\varepsilon_1^t|U(t)|1\rangle\big|^2\mathcal{N}(W|\varepsilon_1^t+\omega_0,\tilde{\sigma}) \\
&+\cos^2\theta\big|\langle\varepsilon_2^t|U(t)|1\rangle\big|^2\mathcal{N}(W|\varepsilon_2^t+\omega_0,\tilde{\sigma}) \\
&-\sin^2\theta\big|\langle\varepsilon_1^t|U(t)|2\rangle\big|^2\mathcal{N}(W|\varepsilon_1^t-\omega_0,\tilde{\sigma}) \\
&-\sin^2\theta\big|\langle\varepsilon_2^t|U(t)|2\rangle\big|^2\mathcal{N}(W|\varepsilon_2^t-\omega_0,\tilde{\sigma})\Big|,
\end{split}
\end{equation}
where $\mathcal{N}(W|\mu,\tilde{\sigma})\equiv\exp\{-(W-\mu)^2/(2\tilde{\sigma}^2)\}/(\sqrt{2\pi}\tilde{\sigma})$ is the normal distribution of $W$ with $\mu$ being the average value and $\tilde{\sigma}=\sqrt{2}\sigma$ being the variance. In Fig. 1 we depict $\mathcal{D}_W$ (black solid curves) as a function of measurement error $\sigma$ for different states at fixed time. For the projective measurement $\sigma=0$, $\mathcal{D}_W=1$, in this case, energy levels before external driving can be completely distinguished by quantum work distribution. As $\sigma$ increases, $\mathcal{D}_W$ is rapidly reduced to $D\equiv|\rho_{11}-\rho_{22}|=\cos2\theta$ (see the green short dashed line in Fig. 1) being the energy level predictability of initial state. It is worth stressing that, the work measurement scheme we consider improves the predictability of energy levels, i.e., $\mathcal{D}_W\geq D$. Even if energy levels are degenerate, i.e., $\omega_0=0$, the predictability of levels after work measurement is $\mathcal{D}_W=\int dW\{\big|\cos^2\theta|\langle\varepsilon_1^t|U(t)|1\rangle|^2-\sin^2\theta|\langle\varepsilon_1^t|U(t)|2\rangle|^2\big|\mathcal{N}(W|\varepsilon_1^t,\tilde{\sigma}) +\big|\cos^2\theta|\langle\varepsilon_2^t|U(t)|1\rangle|^2-\sin^2\theta|\langle\varepsilon_2^t|U(t)|2\rangle|^2\big|\mathcal{N}(W|\varepsilon_2^t,\tilde{\sigma})\}\geq0$, which means that the original indistinguishable energy levels can be distinguished by work distribution after measurement. For a special driving process that $|\langle\varepsilon_1^t|U(t)|1\rangle|^2=|\langle\varepsilon_2^t|U(t)|2\rangle|^2=1$ (or $|\langle\varepsilon_2^t|U(t)|1\rangle|^2=|\langle\varepsilon_1^t|U(t)|2\rangle|^2=1$), the predictability of levels $\mathcal{D}_W$ can be further reduced to $\mathcal{D}_W=\int dW|\cos^2\theta\mathcal{N}(W|\varepsilon_1^t,\tilde{\sigma})-\sin^2\theta\mathcal{N}(W|\varepsilon_2^t,\tilde{\sigma})|$ (or $\mathcal{D}_W=\int dW|\cos^2\theta\mathcal{N}(W|\varepsilon_2^t,\tilde{\sigma})-\sin^2\theta\mathcal{N}(W|\varepsilon_1^t,\tilde{\sigma})|$). In this case, if $\tilde{\sigma}\ll|\varepsilon_1^t-\varepsilon_2^t|$, $\mathcal{D}_W\rightarrow1$ and the original indistinguishable energy levels can be completely distinguished by work distribution.

According to Eq. (\ref{visibility}), the effectiveness of quantum coherence is
\begin{equation}\label{VWE}
\begin{split}
\mathcal{V}_W=2\tilde{C}\Big|\mathrm{Re}[\langle\varepsilon_1^t|U(t)|1\rangle\langle2|U^\dag(t)|\varepsilon_1^t\rangle]\Big|
\mathrm{erf}\Bigl(\frac{\varepsilon^t}{2\sigma}\Bigr),
\end{split}
\end{equation}
where $\tilde{\mathcal{C}}=\mathcal{C}e^{-\omega_0^2/\tilde{\sigma}^2}$ is the survived quantum coherence after the first measurement with $\mathcal{C}\equiv\mathrm{Tr}|\rho_c|=|\sin2\theta|$ being the degree of quantum coherence before the first measurement and $e^{-\omega_0^2/\tilde{\sigma}^2}$ quantifying the degree of the reduction of quantum coherence after the first measurement. If $\omega_0=0$, i.e., the energy levels of the system before driving are degenerate, quantum coherence can not be destroyed. In other words, quantum coherence between the degenerate levels, i.e., the internal coherence \cite{Kwon2018} can not be destroyed by the first measurement. If $\omega_0\neq0$, i.e., energy levels are non-degenerate, quantum coherence (external coherence \cite{Kwon2018}) is inevitably reduced by the first measurement, but the reduction of quantum coherence can be suppressed by measurement error $\sigma$. The lager the error, the less the reduction of quantum coherence, if $\sigma\rightarrow\infty$, $\tilde{\mathcal{C}}\rightarrow\mathcal{C}$. However on the contrary, the manifestation of quantum coherence quantified by $|\mathrm{Re}[\langle\varepsilon_1^t|U(t)|1\rangle\langle2|U^\dag(t)|\varepsilon_1^t\rangle]|
\mathrm{erf}(\frac{\varepsilon^t}{2\sigma})$ is limited by measurement error $\sigma$ (see $\mathrm{erf}(\frac{\varepsilon^t}{2\sigma})$). If $\sigma\rightarrow0$, $\mathrm{erf}[\varepsilon^t/(2\sigma)]\approx1$, the survived quantum coherence (if it exists) can be completely manifested. On the other hand, if $\sigma\rightarrow\infty$, $\mathrm{erf}[\varepsilon^t/(2\sigma)]\approx0$, there is no quantum coherence can be manifested by quantum work distribution. The tradeoff between the destruction and the manifestation of quantum coherence implies that there is an optimal measurement error that maximizes the effectiveness of quantum coherence. In Fig. 1 we depict $\mathcal{V}_W$ (olive dashed curves) as a function of measurement error $\sigma$ for different states at fixed time. It can be seen that when measurement error $\sigma\rightarrow0$ (i.e., projective energy measurement), the effectiveness of quantum coherence $\mathcal{V}_W=0$. As $\sigma$ increases, $\mathcal{V}_W$ firstly increases to its maximum value, the quantum coherence effectiveness of initial state $\mathcal{V}=\mathcal{C}=|\sin2\theta|$ (see the magenta dashed dotted dotted line in Fig. 1) and then reduces to 0. Only for the maximally coherent state $|\Psi\rangle=(|2\rangle\pm|1\rangle)/\sqrt{2}$ ($\theta=\pi/4$), the inequality $\mathcal{V}_W\leq1$ can be saturated (see Fig. 1(c)).

Comparing $\mathcal{D}_W$ and $\mathcal{V}_W$ in Fig. 1 we can see that when $\mathcal{D}_W$ is maximized, $\mathcal{V}_W$ is minimized and vice versa. This mutually exclusive behaviors of $\mathcal{D}_W$ and $\mathcal{V}_W$ further illustrate the fundamental tradeoff relation $\mathcal{D}_W^2+\mathcal{V}_W^2\leq1$ shown in Eq. (\ref{DV2_inequality}). In Fig. 1, we also depict $\mathcal{D}_W^2+\mathcal{V}_W^2$ (red dashed dotted curves) as a function of measurement error $\sigma$ for different states at fixed time. It can be seen that the inequality of $\mathcal{D}_W^2+\mathcal{V}_W^2\leq1$ is always satisfied and its bound is saturated when $\mathcal{D}_W$ is maximized or $\mathcal{V}_W$ is maximized. The tradeoff relation $\mathcal{D}_W^2+\mathcal{V}_W^2\leq1$ implies another equivalent uncertainty relation $\mathcal{D}_W+\mathcal{V}_W\leq\sqrt{2}$. The state that saturates this uncertainty inequality (i.e., $\mathcal{D}_W+\mathcal{V}_W=\sqrt{2}$) is the minimum-uncertainty state we named because the uncertainty of the whole information of population and coherence is minimal (like the minimum-uncertainty state in quantum optics where the uncertainty of coordinate and momentum is minimal and the uncertainty relation $\Delta x \Delta p\geq\hbar/2$ is saturated). In the driven two-level system we consider, the minimal uncertainty state is $|\Psi\rangle=\sin(\pi/8)|2\rangle\pm\cos(\pi/8)|1\rangle$ (see Fig. 1(b)).

\section{Conclusions}
In this paper we mainly investigated the duality in quantum work distribution and gave a fundamental uncertainty relations of  $\mathcal{D}_W^2+\mathcal{V}_W^2\leq1$ between contributions of population and coherence to quantum work distribution, which leads to a deeper understanding of the effects of quantum coherence in quantum thermodynamics. Finally, we considered a driven two-level system and gave the condition of the bound of $\mathcal{D}_W^2+\mathcal{V}_W^2$. Our discussion above only considered the population and coherence of initial state. However, according to $\mathcal{P}(W)=\mathrm{Tr}[\mathcal{M}^W\rho]=\mathrm{Tr}[\mathcal{M}^WU^\dag(t)\rho(t) U(t)]=\mathrm{Tr}[U(t)\mathcal{M}^WU^\dag(t)\rho(t)]$, we can define a new POVM measurement $\{U(t)\mathcal{M}^WU^\dag(t)\}$ and divide the state $\rho(t)$ into coherent and incoherent parts in the eigenbasis of Hamiltonian $H(\lambda_t)$. Then we can investigate the contributions of the population and coherence of the energy levels at any time $t$, and the results are similar to that of initial state. It would also be interesting to apply the perspective of ``wave-particle" duality to investigate the bounds of amount and efficiency of work extraction from quantum coherence, etc.

\section{acknowledgement}
This work was supported by the National Natural Science Foundation of China (Grants No. 11705099, No. 11675017 and No. 11775019).


\begin{thebibliography}{}

%ÄÉÃ×ʵÑé
\bibitem{Bloch2008} I. Bloch, J. Dalibard, and W. Zwerger, Many-body physics with ultracold gases, Rev. Mod. Phys. \textbf{80}, 885 (2008).
\bibitem{Kippenberg2007} T. J. Kippenberg, K. J. Vahala, and J. Kerry, Cavity opto-mechanics, Opt. Express \textbf{15}, 17172 (2007).
\bibitem{Aspelmeyer2014} M. Aspelmeyer, T. J. Kippenberg, and F. Marquardt, Cavity optomechanics, Rev. Mod. Phys. \textbf{86}, 1391 (2014).

\bibitem{Talkner2007a} P. Talkner, E. Lutz, and P. H\"{a}nggi, Fluctuation theorems: Work is not an observable, Phys. Rev. E \textbf{75}, 050102(R) (2007).

%¾­µäÕÇÂä
\bibitem{Jarzynski1997} C. Jarzynski, Nonequilibrium Equality for Free Energy Differences, Phys. Rev. Lett. \textbf{78}, 2690 (1997).
\bibitem{Crooks1999} G. E. Crooks, Entropy production fluctuation theorem and the nonequilibrium work relation for free energy differences, Phys. Rev. E \textbf{60}, 2721 (1999).
\bibitem{Seifert2005} U. Seifert, Entropy Production along a Stochastic Trajectory and an Integral Fluctuation Theorem, Phys. Rev. Lett. \textbf{95}, 040602 (2005).
\bibitem{Esposito2010} M. Esposito, Three Detailed Fluctuation Theorem, Phys. Rev. Lett. \textbf{104}, 090601 (2010).
\bibitem{Ueda2010} T. Sagawa and M. Ueda, Generalized Jarzynski Equality under Nonequilibrium Feedback Control, Phys. Rev. Lett. \textbf{104}, 090602 (2010).
\bibitem{Harris2007} R. J. Harris and G. M. Sch\"{u}tz, Fluctuation theorems for stochastic dynamics, J. Stat. Mech. (2007), P07020.
\bibitem{Seifert2012} U. Seifert, Stochastic thermodynamics, fluctuation theorems and molecular machines, Rep. Prog. Phys. \textbf{75}, 126001 (2012).
\bibitem{Jarzynski2004} C. Jarzynski and D. K. W\'{o}jcik Classical and Quantum Fluctuation Theorems for Heat Exchange, Phys. Rev. Lett. \textbf{92}, 230602 (2004).


%ÕÇÂä¹ØϵµÄÁ¿×ÓÀ©Õ¹
\bibitem{Tasaki2000} H. Tasaki, Jarzynski relations for quantum systems and some applications, arXiv:cond-mat/0009244 (2000).
\bibitem{Kurchan2000} J. Kurchan, A quantum fluctuation theorem, arXiv:condmat/0007360 (2000).
\bibitem{Mukamel2003} S. Mukamel, Quantum Extension of the Jarzynski Relation: Analogy with Stochastic Dephasing, Phys. Rev. Lett. \textbf{90}, 170604 (2003).
\bibitem{Talkner2007b} P. Talkner, and P. H\"{a}nggi, The Tasaki-Crooks quantum fluctuation theorem, J. Phys. A \textbf{40}, F569 (2007).
\bibitem{Talkner2009} P. Talkner, M. Campisi, and P. H\"{a}nggi, Fluctuation theorems in driven open quantum systems, J. Stat. Mech. (2009) P02025.
\bibitem{Campisi2009} M. Campisi, P. Talkner, and P. H\"{a}nggi, Fluctuation Theorem for Arbitrary Open Quantum Systems, Phys. Rev. Lett. \textbf{102}, 210401 (2009).
\bibitem{Andrieux2009} D. Andrieux, P. Gaspard, T. Monnai and S. Tasaki, The fluctuation theorem for currents in open quantum systems, New J. Phys. \textbf{11}, 043014 (2009).
\bibitem{Crooks2008} G. E. Crooks, On the Jarzynski relation for dissipative quantum dynamics, J. Stat. Mech. (2008) P10023.
\bibitem{Watanabe2014} G. Watanabe, B. P. Venkatesh, and P. Talkner, Generalized energy measurements and modified transient quantum fluctuation theorems, Phys. Rev. E \textbf{89}, 052116 (2014).
\bibitem{Esposito2009} M. Esposito, U. Harbola, and S. Mukamel, Nonequilibrium fluctuations, fluctuation theorems, and counting statistics in quantum systems, Rev. Mod. Phys. \textbf{81}, 1665 (2009).
\bibitem{Campisi2011} M. Campisi, P. H\"{a}nggi, and P. Talkner, Quantum fluctuation relations: Foundations and applications, Rev. Mod. Phys. \textbf{83}, 771 (2011).

%Á¿×ÓÕÇÂäµÄʵÑéÑéÖ¤
\bibitem{Batalho2014} T. B. Batalh\~{a}o, A. M. Souza, L. M., R. Auccaise, R. S. Sarthour, I. S. Oliveira, J. Goold, G. De Chiara, M. Paternostro, and R. M. Serra, Experimental Reconstruction of Work Distribution and Study of Fluctuation Relations in a Closed Quantum System, Phys. Rev. Lett. \textbf{113}, 140601 (2014).
\bibitem{Huber2008} G. Huber, F. S.-Kaler, S. Deffner, and E. Lutz, Employing Trapped Cold Ions to Verify the Quantum Jarzynski Equality, Phys. Rev. Lett. \textbf{101}, 070403 (2008).
\bibitem{An2015} S. An, J.-N. Zhang, M. Um, D. Lv, Y. Lu, J. Zhang, Z.-Q. Yin, H. T. Quan, and K. Kim, Experimental test of the quantum Jarzynski equality with a trapped-ion system, Nature Phys. \textbf{11}, 193 (2015).
\bibitem{Hoang2018} T. M. Hoang, R. Pan, J. Ahn, J. Bang, H. T. Quan, and T. Li, Experimental Test of the Differential Fluctuation Theorem and a Generalized Jarzynski Equality for Arbitrary Initial States, Phys. Rev. Lett. \textbf{120}, 080602 (2018).
\bibitem{Xiong2018} T. P. Xiong, L. L. Yan, F. Zhou, K. Rehan, D. F. Liang, L. Chen, W. L. Yang, Z. H. Ma, M. Feng, and V. Vedral, Experimental Verification of a Jarzynski-Related Information-Theoretic Equality by a Single Trapped Ion, Phys. Rev. Lett. \textbf{120}, 010601 (2018).
\bibitem{Cerisola2017} F. Cerisola, Y. Margalit, S. Machluf, A. J. Roncaglia, J. P. Paz, and R. Folman, Using a quantum work meter to test non-equilibrium fluctuation theorems, Nat. Commun. \textbf{8}, 1241 (2017).
\bibitem{Naghiloo2018} M. Naghiloo, J. J. Alonso, A. Romito, E. Lutz, and K. W. Murch, Information Gain and Loss for a Quantum Maxwell¡¯s Demon, Phys. Rev. Lett. \textbf{121}, 030604 (2018).



%°üº¬Ïà¸É
\bibitem{Talkner2016} P. Talkner and P. H\"{a}nggi, Aspects of quantum work, Phys. Rev. E \textbf{93}, 022131 (2016).


\bibitem{Solinas2017} P. Solinas, H. J. D. Miller, J. Anders, Measurement-dependent corrections to work distributions arising from quantum coherences, Phys. Rev. A \textbf{96}, 052115 (2017).
\bibitem{Roncaglia2014} A. J. Roncaglia, F. Cerisola, and J. P. Paz, Work Measurement as a Generalized Quantum Measurement, Phys. Rev. Lett. \textbf{113}, 250601 (2014).
\bibitem{Chiara2015} G. D. Chiara, A. J. Roncaglia and J. P. Paz, Measuring work and heat in ultracold quantum gases, New J. Phys. \textbf{17}, 035004 (2015).
\bibitem{Allahverdyan2014} A. E. Allahverdyan, Nonequilibrium quantum fluctuations of work, Phys. Rev. E \textbf{90}, 032137 (2014).

\bibitem{Solinas2015} P. Solinas and S. Gasparinetti, Full distribution of work done on a quantum system for arbitrary initial states, Phys. Rev. E \textbf{92}, 042150 (2015).
\bibitem{Solinas2016} P. Solinas and S. Gasparinett, Probing quantum interference effects in the work distribution, Phys. Rev. A \textbf{94}, 052103 (2016).
\bibitem{Xu2018} B.-M. Xu, J. Zou, L.-S. Guo, and X.-M. Kong, Effects of quantum coherence on work statistics, Phys. Rev. A \textbf{97}, 052122 (2018).



\bibitem{Sampaio2018} R. Sampaio, S. Suomela, T. Ala-Nissila, J. Anders, and T. G. Philbin, Quantum work in the Bohmian framework, Phys. Rev. A \textbf{97}, 012131 (2018).

\bibitem{Aberg2018} J. {\AA}berg, Fully Quantum Fluctuation Theorems, Phys. Rev. X \textbf{8}, 011019 (2018).
\bibitem{Holmes2018} Z. Holmes, S. Weidt, D. Jennings, J. Anders, and F. Mintert, Coherent fluctuation relations: from the abstract to the concrete, Quantum \textbf{3}, 124 (2019).


\bibitem{Alonso2016} J. J. Alonso, E. Lutz, and A. Romito, Thermodynamics of Weakly Measured Quantum Systems, Phys. Rev. Lett. \textbf{116}, 080403 (2016).





\bibitem{Llobet2017} M. P.-Llobet, E. B\"{a}umer, K. V. Hovhannisyan, M. Huber, and A. Ac\'{i}n, No-Go Theorem for the Characterization of Work Fluctuations in Coherent Quantum Systems, Phys. Rev. Lett. \textbf{118}, 070601 (2017).



%²¨Á£¶þÏóÐÔ
\bibitem{Englert1996} B.-G. Englert, Fringe Visibility and Which-Way Information: An Inequality, Phys. Rev. Lett. \textbf{77}, 2154 (1996).


\bibitem{Durr1998} S. D\"{u}rr, T. Nonn, and G. Rempe, Fringe Visibility and Which-Way Information in an Atom Interferometer, Phys. Rev. Lett. \textbf{81}, 5705 (1998).
\bibitem{Peter1999} P. D. D. Schwindt, P. G. Kwiat, and B.-G. Englert, Quantitative wave-particle duality and nonerasing quantum erasure, Phys. Rev. A \textbf{60}, 4285 (1999).
\bibitem{Jacques2008} V. Jacques, E. Wu, F. Grosshans, F. Treussart, P. Grangier, A. Aspect, and J.-F. Roch, Delayed-Choice Test of Quantum Complementarity with Interfering Single Photons, Phys. Rev. Lett. \textbf{100}, 220402 (2008).
\bibitem{Bera2015} M. N. Bera, T. Qureshi, M. A. Siddiqui, and A. K. Pati, Duality of quantum coherence and path distinguishability, Phys. Rev. A \textbf{92}, 012118 (2015).
\bibitem{Bagan2016} E. Bagan, J. A. Bergou, S. S. Cottrell, and M. Hillery, Relations between Coherence and Path Information, Phys. Rev. Lett. \textbf{116}, 160406 (2016).
\bibitem{Coles2016} P. J. Coles, Entropic framework for wave-particle duality in multipath interferometers, Phys. Rev. A \textbf{93}, 062111 (2016).
\bibitem{Bagan2018} E. Bagan, J. Calsamiglia, J. A. Bergou, and M. Hillery, Duality Games and Operational Duality Relations, Phys. Rev. Lett. \textbf{120}, 050402 (2018).
\bibitem{Qureshi2017} T. Qureshi, M. A. Siddiqui, Wave¨Cparticle duality in N-path interference, Ann. Phys. \textbf{385}, 598-604 (2017).

\bibitem{coherence} T. Baumgratz, M. Cramer, and M. B. Plenio, Quantifying Coherence, Phys. Rev. Lett. \textbf{113}, 140401 (2014).

%ÄÚ²¿Ïà¸ÉºÍÍⲿÏà¸É
\bibitem{Kwon2018} H. Kwon, H. Jeong, D. Jennings, B. Yadin, and M. S. Kim, Clock-Work Trade-Off Relation for Coherence in Quantum Thermodynamics, Phys. Rev. Lett. \textbf{120}, 150602 (2018).


\end{thebibliography}
\end{document}